\documentclass[aps,prb,twocolumn,floatfix,goodfloat,showpacs]{revtex4}
\usepackage{graphics}
\usepackage{epsf}
\usepackage{graphics}
\usepackage{bbm}

\newcommand \be {\begin{equation}}
\newcommand \ee {\end{equation}}
\newcommand \bea {\begin{eqnarray}}
\newcommand \eea {\end{eqnarray}}

\newcommand \bi {\bibitem}

\newcommand \lan {\langle}
\newcommand \ran {\rangle}

\newcommand{\myscalebox}[1]{\scalebox{0.33}[0.33]{#1}}

\begin{document}

\title{Ground-state energy fluctuations in the Sherrington-Kirkpatrick 
model}

\author{Matteo Palassini}

\email{matteo@ipno.in2p3.fr}

\affiliation{ Laboratoire de Physique Th\'eorique et Mod\`eles 
Statistiques\\
B\^at.~100, Universit\'e Paris-Sud, Orsay Cedex 91405, France}

\date{August 26, 2003}

\begin{abstract}
The probability distribution function (PDF) of the ground-state energy in 
the Sherrington-Kirkpatrick spin-glass model is numerically determined 
by collecting a large statistical sample of 
ground states, computed using a genetic algorithm. It is shown that 
the standard deviation of the ground-state energy per spin
scales with the number of spins, $N$, as $N^{-\rho}$ with $\rho 
\simeq 0.765$, but the value $\rho=3/4$ is also compatible with 
the data, while the previously proposed value $\rho=5/6$ is ruled out. 
The PDF satisfies
finite-size scaling with a non-Gaussian asymptotic PDF, which
can be fitted remarkably well by the Gumbel distribution for the 
$m$-th smallest element in a set of random variables, with $m \simeq 6$.

%\maketitle

\end{abstract}

{\rm }
\maketitle

Models with quenched disorder describe a wide class of
systems in physics and other fields  \cite{books,YOUNG}. In thermal
systems, the effects of disorder are especially important at low
temperatures, where the physics is dominated by the low-energy states.
These states are also of direct interest in many applications, for
instance to combinatorial search problems \cite{books}.
Understanding the properties of the low-energy states is
therefore a central task.

In this context,
an important aspect concerns the statistical fluctuations of
the energy of the low-lying states with respect to the disorder.
Consider a system described by an energy function
$H_N({\bf S},{\bf J})$, where ${\bf S}$ represents $N$ degrees of freedom
and  ${\bf J}$ a set of quenched random variables signifying
the disorder. The ground-state energy per degree of freedom,
$e_N \equiv N^{-1} \min_{{\bf S}} H_N({\bf S},{\bf J})$,
when considered as a function of ${\bf J}$, is also random variable,
and typically one would like to know its mean  $\lan e_N \ran$,
standard deviation $\sigma_N=(\lan e_N^2 \ran- \lan e_N \ran^2)^{1/2}$,
and possibly its entire
PDF, $p_N(e_N)$.
In the thermodynamic limit, $N\to\infty$, usually the properties
of  extensivity ($\lan e_N \ran \to e_0$, with $e_0$ a quantity
of order one)
and self-averaging ($\sigma_N \to 0$) hold,
so one simply has $p_N(x)\to \delta (x - e_0)$.
Thus, the interesting question is to determine the finite-$N$ 
scaling behavior. More precisely, one may ask for example:

({\em i}\/) How does $\lan e_N \ran$ scale with $N$?

({\em ii}\/) How does $\sigma_N$ scale with $N$? 

({\em iii}\/) Given the natural 
scaling variable $y_N = (e_N - \lan e_N \ran)/\sigma_N$,
does its PDF $\tilde{p}_N(y_N)$ converge for large $N$ to a 
nontrivial asymptotic PDF, $\tilde{p}_\infty(y_N)$?

To leading order, we expect
$\lan e_N \ran = e_0 + b N^{-\omega}$ and $\sigma_N \sim N^{-\rho}$,
so ({\em i}\/) and ({\em ii}\/) amount to determine 
the exponents $\omega$ and $\rho$. Question 
({\em iii\/}) amounts to determine whether the 
scaling relation
\be
p_N(e_N) =
{1\over \sigma_N} \, \tilde{p}_\infty \left(\frac{e_N - \lan e_N 
\ran}{\sigma_N}\right)
\label{scaling}
\ee
holds for large $N$, and the shape of $\tilde{p}_\infty(x)$.
The answer to ({\em ii}\/) and ({\em iii}\/) is simple for systems with
short-range interactions, where the volume can be subdivided so that
$e_N$ is the sum of $O(N)$ nearly independent contributions. For the central
limit theorem, then, $\sigma_N\sim N^{-1/2}$ (as rigorously proven
for short-range spin glasses \cite{aw})
and $\tilde{p}_\infty(x)$ is a Gaussian for moderate values
of $x$ (we will not be concerned with large deviations here).
For sufficiently long interaction range, in general one has 
$\rho\neq 1/2$ and a non-Gaussian $\tilde{p}_\infty(x)$ \cite{foot_diluted}.

In this paper, the questions above are addressed numerically for
the infinite-range Sherrington-Kirkpatrick
(SK) spin-glass model  \cite{sk}. Much is known about this model,
following Parisi's replica-symmetry-breaking (RSB) 
solution  \cite{rsb,books}, but many issues remain
open, finite-size scaling  being a prominent one.
The exponents $\omega$ and $\rho$ remain unknown analytically, while
previous numerical work indicates 
$\omega \simeq 2/3$  \cite{thesis,krzakala,boettcher}
an $\rho \simeq 3/4$ \cite{cabasino,krzakala}.
Determining the full ground-state energy
PDF is a much more challenging task,
which so far has received limited attention  \cite{krzakala}.
Here, we will address these issues with a much higher statistics than in
previous studies.

The interest of this problem lies also in its relationship \cite{bouchaud}
with {\em extreme value statistics} \cite{gumbel}.
The goal of extreme value  statistics 
is to determine whether the PDF of the $m$-th smallest value
$X$ in a set of random variables $X_1, \dots, X_M$ (the energy levels
in our case) approaches a scaling form
$p_M(X)=1/b_M \, \tilde{p}_\infty[(X-a_M)/b_M]$ for large $M$,
with suitable $a_M, b_M$.
The case of independent, identically distributed ({\em i.i.d.}\/) $X_i$'s
is well understood \cite{gumbel}: if scaling holds,
$\tilde{p}_\infty(x)$  is one of three universal distributions,
depending on the PDF of the $X_i$'s. If the latter
is unbounded and decays faster than a power law
for large $|X_i|$, then $\tilde{p}_\infty(x)$
is the {\em Gumbel}\/ PDF (defined below).
An example of a disordered system with {\em i.i.d.\/} energy levels
is the Random Energy Model  \cite{rem}, but
in general the levels are correlated, and this case
is much less understood (an exception is when $e_N$ is the
sum of {\em spatially}\/ uncorrelated variables, as mentioned above).
The Gumbel form is preserved for short-range correlations
and violated for sufficiently long-range correlations 
\cite{gumbel,bouchaud}.
It is therefore interesting to ask if it applies, at least
approximately, to the SK model,
where full RSB leads to  hierarchically
correlated
energy levels (see, however, Ref.~\onlinecite{dean} for a model
with hierarchical correlations giving a non-universal PDF).

{\em Numerical details --}
We study the standard SK model with energy function
\be
H_N = - \sum_{i < j} J_{ij} S_i S_j
\label{ham}
\ee
where the $S_i$'s are $N$ Ising spins ($S_i=\pm 1$) and the $J_{ij}$'s are 
{\em i.i.d.}\/ random
variables drawn from a Gaussian PDF with zero mean and variance $1/N$.
The sizes studied are $N=19,39,59,99$, and $199$ with,
respectively,
250000, 250000, 180000, 120000, and 64000
independent realizations of the disorder (samples).
The ground state of each sample was determined using a genetic algorithm
\cite{pal,py}.
The $l$-th order cumulants of $p_N(e_N)$, denoted by $k_{l,N}$,
and their statistical errors
were then estimated with a bootstrap method \cite{foot_jack} up to $l=5$.
With large statistics, it is especially important to control
the systematic errors due to occasionally missing the
ground state. This was done by performing much
longer runs for a fraction of the samples, checking
that the change in the cumulants was smaller than
the desired statistical error.

{\em Scaling of the mean and standard deviation --}
Previous numerical studies indicate a value $\omega\simeq 2/3$
for both the SK model  \cite{thesis,krzakala} and diluted
spin glasses  \cite{thesis,krzakala,boettcher,liers}.
The analytical result $\omega=2/3$ was obtained
for the finite-size scaling of the internal energy near
$T_c$ and the Almeida-Thouless line \cite{slanina},
but no result is available deep in the ordered phase.
Our data for $\lan e_N - e_0 \ran N^{2/3}$, using
the exact Parisi result  \cite{rsb} $e_0 = 
 -0.7633\dots$, are shown in the inset of Fig.~1. The
small residual dependence on $N$ \cite{foot_error} 
either indicates that $\omega$ is greater than 2/3 or, if $\omega=2/3$,
that the data are accurate enough to resolve
subleading corrections.
A fit with $\lan e_N \ran = -0.7633 + b N^{-\omega}$ 
gives $\omega=0.673\pm 0.002$  ($\chi^2=11.2$, ${\mbox{ndf}}=3$,
ndf being the number of degrees of freedom), where the
error gives the range for which the goodness-of-fit $Q$ is
larger than $10^{-3}$. A fit with $\omega=2/3$ (leaving $b$ as the only
free parameter) gives $Q<10^{-16}$, which is clearly
not acceptable. Nevertheless, since
corrections of just a few percent 
could accommodate the value $\omega=2/3$ (note the vertical
scale in the inset of Fig.~1), this value
cannot be ruled out. 
Fits with corrections (for example 
$\lan e_N  - e_0 \ran N^{\omega}=b+c N^{-\omega_2}$
with fixed $e_0$ and $\omega$)
are not conclusive since the 
corrections are 
small and thus the parameter range is large.

Turning now to the fluctuations,
in the replica formalism one can write the cumulant
generating function of the (quenched) free-energy as
the annealed free energy of $n$ replicas,
$f(n)$. Crisanti et al. \cite{crisanti} argued that if
the first nonlinear term in the small-$n$ expansion of
$f(n)$ is $n^i$ (in the thermodynamic limit)
then $\rho = 1-1/i$. Kondor  \cite{kondor} found
$f(n) = n f - c n^6$ from the ``truncated model'',
which would imply $\rho=5/6$. 
Recently, Aspelmeier and Moore  \cite{aspel} and
De Dominicis and Di Francesco \cite{dedominicis}
found a higher free-energy (hence better) solution
$f(n) = n f$, from which the conclusion $\rho=5/6$ no longer
follows, and Refs.~\onlinecite{aspel2,krzakala} gave qualitative
arguments indicating $\rho=3/4$.
On the numerical side, Cabasino et al.~\cite{cabasino} found
a result compatible with $\rho=3/4$, but with large statistical errors,
and Bouchaud et al.~\cite{krzakala} found
$\rho\simeq 0.76$, also compatible with $3/4$, but %although
they could not rule out $\rho=5/6$.

\begin{figure}
\begin{center}
  \myscalebox{\includegraphics{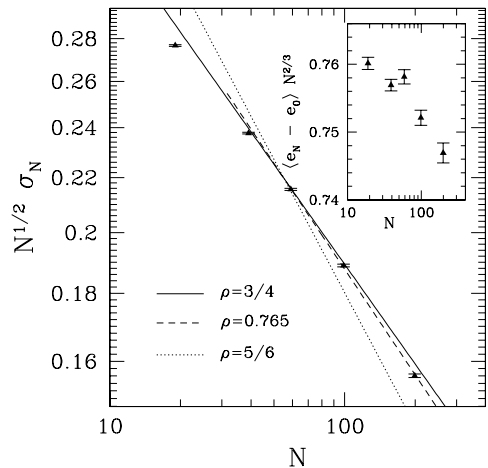}}
\end{center}
\vspace{-0.6cm}
\caption{Standard deviation 
of the ground-state energy per spin as a function
of the system size. The lines represent the power-law $\sigma_N=a 
N^{-\rho}$ for three values of $\rho$.
Inset: finite-size deviation of the mean from 
the RSB result \cite{rsb}
$e_0 =  -0.7633$.}
\label{2cumulant}
\end{figure}

Our numerical data for $\sigma_N$ are displayed in Fig.~1, multiplied
by $N^{1/2}$ to stress that $\sigma_N$ decays faster with $N$
than in short-range models. An exponent  $\rho=3/4$ 
follows reasonably well the data, but deviations 
can be noticed. These are seen more clearly in the plot
of $N^{3/4} \sigma_N$ in Fig.~2(a).
Here the curvature in the data demonstrates the presence of
subleading corrections. 
The dashed line represents a
power law fit $\sigma_N = a N^{-\rho}$ for $N\geq 59$, which
gives $\rho = 0.765 \pm 0.01$ ($\chi^2=3.93, {\mbox{ndf}}=1$),
in agreement with $\rho=3/4$ (the error corresponds again to 
$Q>10^{-3}$).
However, the decreasing trend  for large $N$ \cite{foot_error}
{\em and}\/ the negative curvature suggest 
that $\rho$ may in fact be greater than $3/4$.
As for the mean, fits with correction terms (for fixed $\rho$)
are not very illuminating.
Fig.~1 also shows that $\rho=5/6$ clearly does not fit
the data, as seen also in
Fig.~2(b) where $N^{5/6} \sigma_N$ is far from 
constant. Since the sizes considered are
already well in the scaling regime
(at least as far as Eq.(\ref{scaling}) is concerned, see below),
crossover to a constant for larger $N$ is rather
unlikely, hence we conclude that $\rho=5/6$ is ruled out.
Finally, if  $\sigma_N \sim N^{-\rho}$ and 
Eq.(\ref{scaling}) holds, then $k_{l,N}\sim N^{-l \rho}$.
Figs.~2(c) and 2(d) show that also for $l=3,4$ the data
agree with $\rho\simeq 3/4$ but not with $\rho=5/6$
(note that the statistical error is much larger
than that of $\sigma_N$).

\begin{figure}
\begin{center}
  \myscalebox{\includegraphics{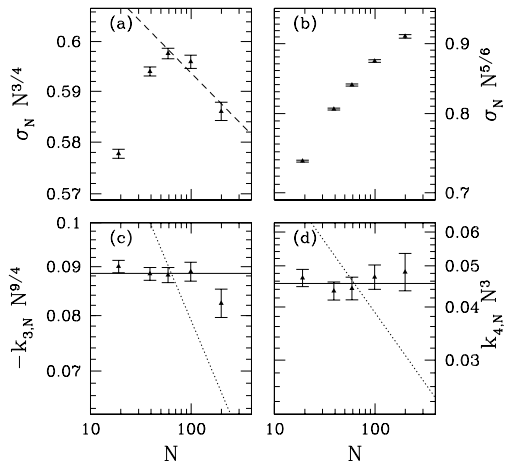}}
\end{center}
\vspace{-0.6cm}
\caption{Plots of $\sigma_N N^{\rho}$ for $\rho=3/4$ (a) and $\rho=5/6$ (b),
and of $k_{l,N} N^{l\rho}$ for $\rho=3/4$, $l=3$ (c) and 
$\rho=3/4$, $l=4$ (d).
The dashed line in (a) is the best fit $\sigma_N=a N^{-\rho}$
for $N\geq 59$, giving $\rho=0.765$.
The dotted lines in (c) and (d) represent power laws
$N^{-l \rho}$ with $\rho=5/6$.}
\label{allcumulant}
\end{figure}

{\em Scaling of the PDF --}
Scaling plots of $p_N(e_N)$ according to Eq.(\ref{scaling}) are shown in 
Fig.~3
and, on a logarithmic scale, Fig.~4.
The data collapse is remarkably good, indicating that we
are already well in the scaling regime.
Comparison with a Gaussian PDF, represented by the dashed
lines in Figs.~3 and 4, clearly indicates a non-Gaussian behavior.
Table I displays the scaled cumulants
$\tilde{k}_{l,N} \equiv k_{l,N}/\sigma_N$ for $l=3,4,5$
(since $\tilde{k}_{2,N}=1$ by definition,
$\tilde{k}_{3,N}$ is the skewness and $\tilde{k}_{4,N}$ the kurtosis
of $\tilde{p}_N$):
the cumulants are independent of $N$ within the errors,
again strongly supporting scaling (small deviations
exist for $N=19$ and $l=3,4$, presumably due to corrections to scaling).
Note that scaling of the PDF
 is independent of the values of $\omega$ and $\rho$.

Next, we compare the scaled PDF to known theoretical
distributions.  The Gumbel PDF for the $m$-th smallest
in a set of {\em i.i.d.} random 
variables has the form \cite{gumbel,foot_gumbel}
\be
g_m(x) = w \exp \left[ m {x-u\over v} -
m \exp {x-u \over v} \right] 
\label{gumbel}
\ee
where $u$ and $v$ are rescaling parameters and $w$ a
normalization constant. To compare this to our data, we impose
zero mean and unit variance which, together with
normalization, fixes $u$, $v$, and $w$
as a function of $m$.
Expressions for the cumulants of $g_m(x)$ are easily obtained. These
are decreasing functions of
$m$, in absolute value, and vanish for large $m$,
thus $g_m$ provides an interpolation
between the standard Gumbel ($m=1$) and a Gaussian PDF ($m=\infty$).
As shown in Table I, the cumulants are fitted well for
$m=6\pm 0.3$ ($\tilde{k}_{5,N}$ shows small deviations 
from the $N\leq 39$ data, possibly an effect of 
corrections to scaling).
Table I also shows that $m=\pi/2$ (see below) is far
from agreeing with the  data, hence even more so is
$m=1$, as also noted in Ref.~\onlinecite{krzakala}.
This shows that the correlations in the energy levels give rise to a
more Gaussian-like PDF with respect to the uncorrelated ($m=1$) case.
The PDF $g_6(x)$ is displayed with solid lines in Figs.~3 and 4,
and the corresponding cumulative distribution in the inset of Fig.~3:
the agreement with the data is excellent
everywhere. Although it is not clear why
a Gumbel distribution with $m\neq 1$ should work here,
$g_6(x)$ can certainly be used to accurately represent the true
PDF up to several standard deviations. It would be interesting to
reach higher statistics to check whether the agreement breaks down.
Incidentally, the Gumbel distribution with a non-integer value
$m=\pi/2$ has been conjectured to describe \cite{bramwell}, at least to
a good approximation, the PDF of spatially averaged quantities in many
correlated systems, even when no extreme
value is apparently involved. A mo\-dified Gumbel form with non-integer $m$
was also used to  describe the order-parameter PDF in three-dimensional
spin glasses \cite{berg}.

Another example of extreme value of correlated random variables is
the smallest eigenvalue, $\lambda_M$, of an $M \times M$ random matrix.
For three well-known ensembles, the scaling PDF of $\lambda_M$
is the Tracy-Widom (TW) distribution \cite{tw}, $f_\beta(x)$, with
$\beta$ a parameter dependent on the ensemble.
We tested it against our data 
finding a poor agreement,
although better than for $g_1(x)$.
Table I shows that the cumulants for $\beta=1$ (and thus 
for $\beta=2$), computed
from the tabulated  $f_\beta(x)$ \cite{spohn}, are
well outside the statistical errors. The PDF $f_1(x)$ is displayed
with dotted lines in Figs.~3 and 4: one can see small deviations
from the data for small $x$ and large deviations in the tails.

In conclusion, we presented accurate numerical data
consistent with $\omega=2/3$ and $\rho=3/4$ when considering
subleading corrections, although 
both exponents may be in fact slightly larger than these values.
The PDF of the ground-state energy satisfies 
finite-size scaling with an asymptotic PDF
empirically well described by a Gumbel distribution.

After this work was virtually finished, we learned that
$p_N(e_N)$ was also studied numerically in Ref.~\onlinecite{andreanov}.

\begin{figure}
\begin{center}
  \myscalebox{\includegraphics{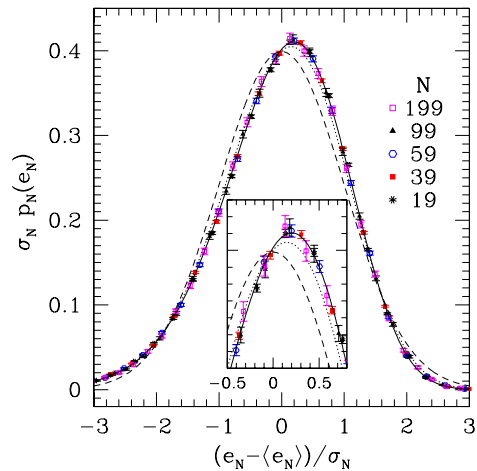}}
\end{center}
\vspace{-0.6cm}
\caption{Scaling plot of the ground-state energy PDF.
The solid line is the Gumbel PDF
with $m=6$ ($u=0.2011219, v=2.348408, w=165.5589$).
The dotted line is the Tracy-Widom PDF with $\beta=1$. 
The dashed line is a Gaussian PDF. All
PDF's are normalized to one and have zero
mean and unit variance. The inset shows a zoom near the origin.}
\label{histo}
\end{figure}

\begin{figure}
\begin{center}
  \myscalebox{\includegraphics{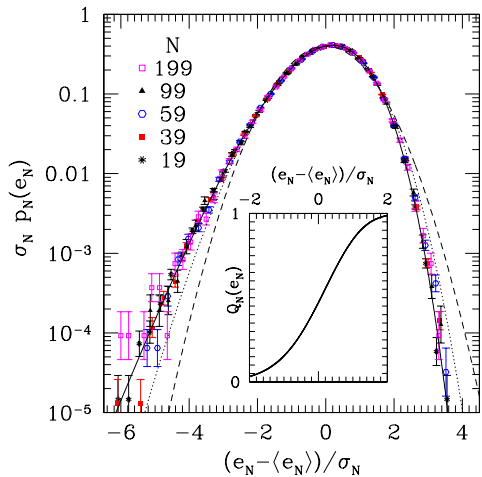}}
\end{center}
\vspace{-0.6cm}
\caption{Same as Fig.~\ref{histo} but on a semi-logarithmic scale. Inset:
scaling plot of the cumulative distribution 
$Q_N(e_N)=\int_{-\infty}^{e_N} p_N(x)\,dx$
for $N=19,39,59,99,199$,
and of the Gumbel cumulative distribution for $m=6$. The curves all fall
on top of each other within the line thickness.}
\label{histolog}
\end{figure}

\begin{table}[h]
\begin{center}
\begin{tabular*}{\columnwidth}{@{\extracolsep{\fill}}rllll}
\hline
\hline
N & $-\tilde{k}_{3,N}$ & $\tilde{k}_{4,N}$ & $-\tilde{k}_{5,N}$ \\
\hline
19 & $0.467 \pm 0.007$ & $0.42 \pm 0.02$ & $0.46 \pm 0.07$ \\
39 & $0.422 \pm 0.007$ & $0.35 \pm 0.02$ & $0.31 \pm 0.05$ \\
59 & $0.413 \pm 0.008$ & $0.35 \pm 0.02$ & $0.32 \pm 0.07$ \\
99 & $0.420 \pm 0.010$ & $0.37 \pm 0.03$ & $0.40 \pm 0.08$ \\
199& $0.409 \pm 0.015$ & $0.41 \pm 0.04$ & $0.68 \pm 0.15$ \\
\hline
$g_6$ & 0.42468 & 0.35346 & 0.4441   \\
$g_{\pi/2}$ & 0.89373 & 1.53674 & 3.8403  \\
$f_1$ & 0.2935 & 0.16524 & 0.1024  \\
$f_2$ & 0.2241 & 0.09345 & 0.0386 \\
\hline
\hline
\end{tabular*}
\end{center}
\caption{Cumulants of the scaled PDF $\tilde{p}_N(y_N)$
for various values of $N$,
compared with the cumulants of the theoretical PDF's $g_m(x)$
(Gumbel distribution for $m$-th largest extreme) and $f_\beta(x)$
(Tracy-Widom distribution).}
\label{tab_samples}
\end{table}

{\em Acknowledgements --}
The author thanks A.P.~Young and M.A.~Moore for very useful exchanges,
and the authors of Ref.~\onlinecite{andreanov} for communicating some of
their findings prior to publication. The author also thanks M.~M\'ezard
for a useful discussion and M.~Pr\"ahofer and H.~Spohn
for making their computation of the TW distribution publicly available
\cite{spohn}.
Support from STIPCO, EC grant number HPRN-CT-2002-00319,
is acknowledged.


\vspace{-0.6cm}
\begin{thebibliography}{99}

\bi{books} M.~M\'ezard, G.~Parisi, and M.A.~Virasoro, {\em Spin Glass
Theory and Beyond} (World Scientific, Singapore, 1987).

\bi{YOUNG} {\em Spin Glasses and Random Fields}, A.P.~Young Ed.
(World Scientific, Singapore, 1998).

\bi{aw} J.~Wehr and M.~Aizenmann, J. Stat. Phys. {\bf 60}, 287 (1990).

\bi{foot_diluted}
Interestingly, infinite-range spin glasses on finite-connectivity graphs
seem to behave like the short-range case \cite{krzakala}, although the same
argument cannot be applied.

\bi{sk} D.~Sherrington and S.~Kirkpatrick, Phys. Rev. Lett. {\bf 35}, 1792 
(1975).

\bi{rsb} G.~Parisi, Phys. Rev. Lett. {\bf 43}, 1754 (1979); {\bf 50}, 1946 
(1983).

\bi{thesis} M.~Palassini, PhD thesis, 2000.

 \bi{krzakala}
J.-P.~Bouchaud, F.~Krzakala, and O.C.~Martin, cond-mat/0212070.

\bi{boettcher} S.~Boettcher, Eur. Phys. J. B {\bf 31}, 29 (2003).

\bi{cabasino}
S.~Cabasino, E.~Marinari, P.~Paolucci, and G.~Parisi,
J. Phys. A {\bf 21}, 4201 (1988).

\bi{bouchaud} J.-P.~Bouchaud and M.~M\'ezard, J. Phys. A {\bf 30}, 7997  
(1997).

\bi{gumbel} E.J.~Gumbel, {\em Statistics of Extremes} (Columbia University 
Press, New York, 1958); 
J.~Galambos, {\em The Asymptotic Theory of Extreme Order 
Statistics}
(R.E. Krieger Publishing Co., Malabar, FL, 1987).


\bi{rem} B.~Derrida, Phys. Rev. B {\bf 24}, 2613 (1981); D.J.~Gross and
M.~M\'ezard, Nucl. Phys. B {\bf 240}, 431 (1984).

\bi{dean}
D.S.~Dean and S.N.~Majumdar, Phys. Rev. E. {\bf 64}, 046121 (2001).

\bi{pal}  K.~F.~Pal, Physica A {\bf 223},  283 (1996).

\bi{py}
M.~Palassini and A.P.~Young,  Phys. Rev. Lett. {\bf 83}, 5126 (1999).

\bi{foot_jack} A jackknife method was also tried, but this performed poorly
for large-order cumulants.

\bi{liers}
F.~Liers, M.~Palassini, A.K.~Hartmann, and M.~Juenger, cond-mat/0212630.


\bi{slanina} G.~Parisi, F.~Ritort, and F.~Slanina,
J. Phys. A {\bf 26}, 247 (1993); J. Phys. A {\bf 26}, 3775  (1993).

\bi{foot_error}
The decreasing trend for large $N$ in the inset of Fig.~1 and in 
Fig.~2(a) is {\em not} due to systematic errors from occasionally
missing the ground state
at large $N$, since these would give {\em higher}\/ values
of both $\langle e_N \rangle$ and $\sigma_N$ (this was 
also verified numerically by performing shorter runs).


\bi{crisanti}
A.~Crisanti, G.~Paladin, H.-J.~Sommers, and A.~Vulpiani, J. Phys. I France
{\bf 2}, 1325 (1992).

\bi{kondor} I.~Kondor, J. Phys. A {\bf 16}, L127 (1983).

\bi{aspel}
T.~Aspelmeier and M.A.~Moore, Phys. Rev. Lett. {\bf 90}, 177201 (2003).

\bi{dedominicis} C.~De Dominicis and P.~Di Francesco, cond-mat/0301066.

\bi{aspel2}
T.~Aspelmeier, M.A.~Moore, and A.P.~Young, Phys. Rev. Lett. {\bf 90}, 127202 
(2003).


\bi{foot_gumbel} The case $u=0, v=1$ is properly
known as the Gumbel distribution,
while the more general case is often called Fisher-Tippett distribution.

\bi{bramwell} S.T.~Bramwell et al., Phys. Rev. Lett. {\bf 84}, 3744 (2000);
Phys. Rev. E {\bf 63}, 041106 (2001).

\bi{berg}
B.A.~Berg, A.~Billoire, and W.~Janke, Phys. Rev. E {\bf 65}, 045102R (2002).

\bi{tw}
C.A. Tracy and H. Widom, Comm. Math. Phys. {\bf 159}, 151 (1994);
{\bf 177}, 727 (1996).

\bibitem{spohn} M. Pr\"ahofer and H. Spohn,  Physica A {\bf 279}, 342 
(2000);
Phys. Rev. Lett. {\bf 84}, 4882 (2000); J. Stat. Phys. {\bf 108}, 1071 
(2002); Data available at {\tt http://www-m5.mathematik.tu-muenchen.de/KPZ/}.

\bi{andreanov} A.~Andreanov, F.~Barbieri, and O.C.~Martin, cond-mat/0307709.


\end{thebibliography}
\end{document}